\documentclass[useAMS]{mn2e}

\title[Infrared studies of  V5558 Sgr: a slow nova  with multiple outbursts]{Near-infrared studies of  V5558 Sgr: an unusually slow nova  with multiple outbursts }

\author[Das et al.]{Ramkrishna Das$^{1}$ \thanks{E-mail: ramkrishna.das@bose.res.in},
Dipankar P. K. Banerjee$^{2}$ \thanks{E-mail: orion@prl.res.in}, Arpita Nandi$^{1}$, N M Ashok$^{2}$ \and $\&$ Soumen Mondal$^{1}$\\
\\
$^{1}$Astrophysics \& Cosmology Department, S N Bose National Centre for Basic Sciences, Salt Lake, Kolkata 700098, India.\\
$^{2}$Astronomy \& Astrophysics Division, Physical Research Laboratory, Navrangpura, Ahmedabad 380009, India.\\}

\usepackage{graphicx}
\usepackage{footnote}
\usepackage{threeparttable}

\begin{document}

\maketitle

\label{firstpage}

\begin{abstract}

We present near-infrared (1-2.5 $\mu$m) $JHK$ photo-spectroscopic results of the unusually slow nova V5558 Sgr (2007). V5558 Sgr showed a slow climb to maximum  that lasted for about 60 days and then underwent at least five strong
secondary outbursts. We have analyzed the optical light curve to derive
large t${_2}$ and t${_3}$ values of 281 $\pm$ 3 and 473 $\pm$ 3 days respectively.
An alternate approach is adopted to derive a distance estimate of 1.55 $\pm$ 0.25 kpc as conventional MMRD relation may not be
applicable for a slow nova.
In the pre-maxima stage the  spectra showed narrow (FWHM $\sim$ 400 - 550 km s$^{-1}$) and strong emission lines of
Paschen and Brackett series with prominent P-Cygni components. In the later phase the spectra show significant
changes with the development of strong and broad ($\sim$ 1000 km s$^{-1}$) emission lines of HI, HeI, OI,
and NI and some uncommon Fe II emission lines. No evidence of dust formation is seen.
V5558 Sgr has been shown to be a rare hybrid nova showing a transition
from He/N to Fe II type from optical spectra.
However the near-infrared data do not show such a transition and we discuss this anomalous
behavior. A recombination analysis of the  Brackett
lines allows us to constrain the electron density and
emission measure during the early optically thick phase and to
estimate the mass of the ejecta to be (6.0 $\pm$ 1.5) $\times$ 10$^{-4}$
M$_{\odot}$,  assuming a filling factor of unity,  from later observations.

\end{abstract}



\section{Introduction}

Nova V5558 Sgr was discovered by Y. Sakurai on 2007 April 14.777 UT (JD 2454205.277) at magnitude 10.3 (Nakano et al. 2007) during its pre-maximum
stage. Henden $\&$ Munari (2007) have reported that no visible progenitor was detected in the POSS plates  suggesting a value of 13.4 magnitudes as the lower limit of the outburst  amplitude. They  derived the astrometric position of the nova to be $\alpha$ = 18$^{h}$10$^{m}$18.258$^{s}$ $\pm$ 0.046$^{s}$,
$\delta$ = -18$^{\circ}$46$^{\prime}$51.95$^{\prime\prime}$ $\pm$ 0.047$^{\prime\prime}$ which is close to the coordinates reported
by Sakurai  (Nakano et al. 2007).
After the discovery, the brightness of the nova increased very slowly.
The maximum brightness was attained on July 10.0 UT at $V$ = 6.53 (Munari et al., 2007b) which was followed by a flattening of the light curve and then a slow decline during which  the nova underwent
four more  re-brightenings (Tanaka et al. 2011). Such a slow rise to maximum accompanied by multiple outbursts during the evolution  have been rarely observed in other novae (for example, V723
Cas (Evans et al. 2003) and HR Del (Terzan 1970)).\\

 Optical studies of V5558 Sgr have been performed by several observers (Iijima 2007a, 2007b; Naito 2007; Munari
et al. 2007a; Poggiani 2008). The spectra recorded during the pre-maxima phase were dominated by the strong Balmer and He I
lines along with weak features of emission lines of Fe II multiplets, Mg II, N II, Si II, Ca I and
[O I]. All of these emission lines exhibited narrow and sharp profiles with typical full widths at half maxima (FWHM)
 of about 480-540 km s$^{-1}$, as measured from Balmer lines, which is a value range  generally associated with Fe II type novae.
Most of the strong emission lines were accompanied by absorption components of  P-Cygni profiles, blueshifted by 400-500 km s$^{-1}$.
Munari et al. (2007b) estimated a reddening of $E(B - V)$ = 0.36 from the Na I D1 and D2 lines.
No significant change in the spectra was noticed during the pre-maximum evolution except the disappearance of He I lines and the emergence of Fe II multiplets in
emission  on 2007 May 11 (Munari et al. 2007a).  \\

Significant changes in the spectra were noticed during and after the initial maximum. The spectra, observed during and a short time after the initial
maxima on 2007 July 10, were characterized by a smooth underlying continuum with numerous weak absorptions along with strong emission lines of Balmer series and
FeII multiplets (Iijima 2007b, Munari et al. 2007b). Also, the He I lines which had disappeared on 2007 May 11, reappeared on 2007 July 12.
Also a remarkable change was noticed in the line profile widths. The emission lines which were narrow in the pre-maxima stage, become strong and
broad (FWHM $\sim$ 1150 km s$^{-1}$, Munari 2007b)
which indicated the presence  of fast hot winds.
The nature of the spectra did not change much during the remaining  part of the evolution till 2007 Nov 2 (Tanaka et al. 2011);
the spectra were dominated by strong emission features of H, Fe II and He I lines along with weak emission features
of OI, NII, CII. Many of these emission
features were accompanied by sharp P-Cygni profile which disappeared gradually with time.
Further spectroscopic follow up during the declining phase by Poggiani (2010) reported the appearance of
forbidden lines, e.g., [OIII] 4959, 5007, and [Fe VI], [Fe VII], [Ca V] which suggests that V5558 Sgr had entered the nebular phase.
Overall, the spectral behavior of V5558 Sgr before and after the maximum, is very similar to that of V723 Cas (Iijima
2007b; Poggiani, 2008, 2010). Poggiani (2010) estimated the characteristic time t$_{3}$ (time taken to decline by 3 magnitudes from visual maximum) = 170 $\pm$ 2 days
making the nova a slow one according to the definition of speed class (Payne-Gaposchin 1957). Using the value of t$_{3}$ Poggiani (2010) revised
the estimated value of the distance to be 1.3 - 1.6 kpc and absolute magnitude (M$_{V}$) at maximum  in the range of 6.3 - 5.9 in agreement
with the value of 6 $\pm$ 1 observed in case of slow novae V723 Cas and HR Del. Using these values Poggiani (2010) derived the white dwarf mass
in the range of 0.58 - 0.63 M$_{\odot}$. We have certain reservations on Poggiani's (2010) method of deriving the distance which is discussed in section 3.2\\

The first infrared observations of V5558 Sgr were formally reported by Lynch et al. (2007).
The infrared spectrum (0.8 - 2.5 $\mu$m) observed on 2007 September 12,
showed strong and narrow (FWHM $\sim$ 620 km s$^{-1}$) emission lines of H I, He I, OI, Ca II, Fe II, and [N I]. All of the H I and
few of the He I lines and Fe II lines showed P-Cygni absorption profiles which suggested significant optical depths in these lines.
Further infrared observations performed by Rudy et al. (2007) on 2007 October 10 and 11 showed a spectrum (0.8 - 5.5 $\mu$m) dominated
by low excitation broad (FWHM $\sim$ 1600 km s$^{-1}$) emission lines with strong N I and Fe II lines. The P-Cygni profiles had
disappeared from the H I lines but were weakly present on the He I lines. They measured a reddening of about 0.8 from the O I
lines. Both Infrared observations found no evidence of dust formation.\\

In this paper we present results of near-infrared $JHK$ (1 - 2.5 $\mu$m) spectroscopic and photometric observations of V5558 Sgr.
The  observations are is described in Section 2 and the results in Section 3.\\

\section{Observations}
V5558 Sgr was observed in the near-Infrared $JHK$ (1-2.5 $\mu$m) region using the 1.2m
telescope at Mount Abu Infrared Observatory operated by Physical Research Laboratory, India. Observations were taken on 7 epochs in the pre-maxima
rising phase and on 1 epoch in the post-maxima phase. The nova could not be followed in the months of mid-June to mid-October due to monsoon season.
The near-infrared $JHK$ spectra presented in this paper were obtained in each of $J, H, K$ bands at  similar dispersions of $\sim$ 9.75 {\AA}/pixel using the Near Infrared Imager/Spectrometer with a 256$\times$256 HgCdTe NICMOS3 array. Each time a set
of at least two spectra of the object at two dithered position along the slit were taken. In the same manner, a set of two spectra of a nearby comparison
star (SAO 161564, spectral type B9.2/A0V, $V$ = 5.126) at similar airmass as the object were taken at
the same grating position in order to remove the atmospheric absorption effects from the nova spectra.
Each set of two images were subtracted from each other to eliminate the background counts which comprises of emission from sky and dark counts from the detector.
From these subtracted images spectra were extracted using IRAF tasks. The spectra were wavelength calibrated using a combination of  OH sky lines and telluric
lines that register with the stellar spectra. Following the standard procedure, Hydrogen Paschen and Brackett absorption lines were removed  from the
spectra of the comparison star and the  nova spectra were then divided by the  spectra of the comparison star. The ratioed spectra were multiplied by a blackbody curve
generated at the effective temperature of the comparison star to get the final spectra. However, caution should be taken while considering any spectral feature
around 1.12 $\mu$m in the $J$ band and between 2 to 2.05 $\mu$m in the K band, especially, in the early phase spectra when the emission lines are not so strong.
This is because some differences in airmass  between the nova and comparison star remained. Hence, the ratioing process, while removing telluric features,  leaves some residuals in these wavelength regions where telluric absorption due to atmospheric oxygen and carbon-dioxide respectively are strong. The log of spectroscopic observations are provided in Table 1.\\

Near-Infrared photometric observations of the nova in the $JHK$ bands were done in photometric sky conditions using the imaging mode
of the NICMOS3 array. Several frames, both of the nova and a selected standard star (SAO 186544, spectral type B0.5Ib/IIC, $V$ = 6.02) in each of the
$J,H,K$ filters, were recorded in 5 dithered positions offset typically by
20 arcsec. Near-infrared $JHK$ magnitudes were then extracted using APPHOT package in IRAF
following the regular procedure for photometric reduction (e.g. Banerjee $\&$ Ashok, 2002) followed by us. The log of the photometric observations and the derived $JHK$ magnitudes  are given in Table 2.
Further observations  beyond November could not be carried out as the object had begun to  approach solar conjunction.\\

\begin{table}
\centering
\caption{Observational log of the spectroscopic observations of nova V5558 Sgr. The date of outburst
is taken to be 2007 Apr 14.777 UT.}\

\begin{tabular}{lcccccc}
\hline\
Date & Days        & &         & Integration time &   \\
2007 & since        & &        & (sec)            &    \\
(UT) & Outburst  & & \emph{J} & \emph{H}         & \emph{K}    \\
\hline
\hline
Apr. 26.951  & 12.174   & & 90      & 90               & 40 \\
Apr. 29.836  & 15.059   & & 30      & 30               & 75 \\
May. 01.861  & 17.084   & & 60      & 60               & 40 \\
May. 05.889  & 21.112   & & 60      & 60               & 40 \\
May. 07.832  & 23.055   & & 30      & 30               & 90 \\
Jun. 07.813  & 54.036   & & 90      & 75               & 45 \\
Jun. 10.844  & 57.067   & & 75      & 75               & 40 \\
Oct 20.594   & 188.817  & & 60      & 60               & 90 \\

\hline

\end{tabular}
\end{table}

\begin{table*}
\centering
\caption{Observational log of the photometry. The date of outburst commencement
is taken to be 2007 Apr 14.777 UT.}\

\begin{tabular}{lccccc}

\hline
Date          & Days         & &             & Magnitudes&          \\
2007          & since        & &             &           &          \\
(UT)          & Outburst     & & \emph{J}    & \emph{H}  & \emph{K} \\
\hline
\hline
Apr. 29.891   & 15.114       & & 7.02 $\pm$ 0.05      & 6.57 $\pm$ 0.09    & 6.09 $\pm$ 0.05 \\
May  01.955   & 17.178       & & 6.94 $\pm$ 0.06      & 6.71 $\pm$ 0.10    & 6.03 $\pm$ 0.10 \\
May 05.963    & 21.186       & & 6.77 $\pm$ 0.11      & 6.35 $\pm$ 0.07    & 5.93 $\pm$ 0.13 \\
May 07.944    & 23.167       & & 6.70 $\pm$ 0.09      & 6.33 $\pm$ 0.05    & 5.89 $\pm$ 0.07 \\
Jun. 07.860   & 54.083       & & 6.29 $\pm$ 0.12      & 5.80 $\pm$ 0.08    & 5.59 $\pm$ 0.12 \\
Jun. 10.798   & 57.021       & & 6.25 $\pm$ 0.10      & 5.84 $\pm$ 0.12    & 5.48 $\pm$ 0.09 \\

\hline
\end{tabular}
\end{table*}

\section{Results}
\subsection{Optical and near-infrared lightcurves}

The partial $V$ band light curve for the first 250 days, to specially highlight the secondary maxima, is shown in Figure 1 while the entire  lightcurve
over an extended period of 6.25 years from $\textbf{maxima}$ 
is shown in Figure 2. In Figure 1 we also show the near-infrared $JHK$ light curves.
The optical light curve is based on the data collected from American
Association of Variable Star Observers (AAVSO) and from Association Francaise des Observateurs dEtoiles Variables (AFOEV). The near-infrared $JHK$
magnitudes are from Mt. Abu observations (see Table 2) and SMARTS/CTIO 1.5m telescope facility (Walter et al., 2012). The epochs of Mount Abu
Infrared observations are marked by vertical dashes on the light curve. After the
outburst the light curve rises very slowly. We find from AAVSO data that the nova took about sixty days to reach $V$ = 8.5 from the discovery magnitude 10.3.
After staying at $V$ = 8.5 , for a few days, the nova underwent its first episode of rebrightening to reach m$_{v}$ = 6.48 on 2007 July 10.56 (JD = 2454292.06) which is in line with the findings of
Munari et al. (2007b). Following this, the brightness of V5558 Sgr started to decrease. During the declining phase, V5558 Sgr showed at
least four more  secondary outbursts occurring on  July 25 ($V$ = 7.7, t = 101.6 days), August 12 ($V$ = 7.0, t = 119 days), September 25
($V$ = 7.1, t = 163 days) and October 31 ($V$ = 7.4, t = 199.6 days) respectively, where t = days after commencement of the eruption. After the fifth maxima V5558 Sgr declined slowly (see Figure 2). The near-infrared light curves are seen to mimic the behavior of the optical lightcurve and do not warrant any special comments. During the pre-maximum stage a gradual rising is seen in all the three
near-infrared $JHK$ bands similar to optical light curve, the J-H and H-K values were about 0.2 and 0.3 respectively.
We do not have any near-infrared observation during the first maximum. However, the SMART/CTIO observations during the other secondary maxima show that
the J-H and H-K have been about 1.1 and in the range of 0.6 - 1.0, respectively. From near-infrared observations we do not notice any excess in any of the near-infrared $JHK$ light curves which indicates that no dust has been formed in the nova ejecta.\\

The light curve shows all the characteristics of the J class of novae
 based on the classification system of Strope et al. (2010).
J class of nova show 'jittering' or 'flare-ups' of substantial amplitude which
could be thought of as outbursts above a smoothed light curve that follows the universal decline law
predicted by Hachisu $\&$ Kato (2006). Strope et al. (2010) listed 14 novae that belong to J type. While we compare the light curve of V5558 Sgr
with these novae, we find close similarities with V723 Cas and HR Del out of these 14 novae. Both of V723 Cas and HR Del showed a
prolonged pre-maximum halts, V723 Cas took 100 days and HR Del took 180 days respectively to reach the
maximum. Both of these novae showed multiple outbursts, V723 Cas showed 6 outbursts in between 100 and 425 days
and HR Del showed 5 outbursts in between 180 and 380 days.\\

\begin{figure}
\centering
\includegraphics[bb= 0 0 340 275, width=3.5in,height=2.7in,clip]{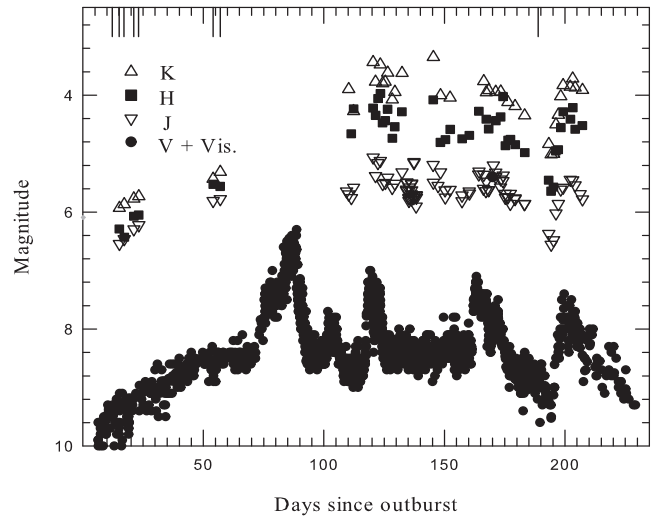}
\caption[]{A magnified view of the secondary outbursts in the  optical  and $JHK$ light curves (black circles and gray crosses respectively) observed in  V5558 Sgr. Optical
data (V band and Visual) are taken from AAVSO and AFOEV databases. The near-Infrared $JHK$ magnitudes are from the 1.2m
Mount Abu Telescope (see Table 2) and from Walter et al. (2012).  The errors of the optical and NIR data are typically smaller than 0.05 mag;  smaller than their respective symbol sizes used.}
\label{fig1}
\end{figure}


The cause behind the occurrence of multi-outbursts, as observed in this nova, is not well understood. Previous studies of slow novae have shown  that multi-outbursts might take place on white dwarfs of mass $\sim$ 0.6 M$_{\odot}$ which is close to the critical value required  beginning of the  thermonuclear runaway
event (for example, Friedjung et al. 1992; Kato
et al. 2002). However, recent calculations with  OPAL opacities, show that thermonuclear
runaway may occur even in much less massive white dwarfs (WDs), e.g., 0.2 - 0.5 M$\odot$ WDs (Shen et al. 2009).
In a recent study, Kato $\&$ Hachisu (2009, 2011) pointed out that two different types of evolution of slow novae of a certain range of
WD masses, $\sim$ 0.5 - 0.7 M$\odot$ may take place. One is the evolution with optically thick winds and the other is without optically thick winds.
The wind-type novae show a sharp peak in the optical light curve
and the light curve decays quickly. On the other hand novae without an  optically thick wind
evolve very slowly and stay at an extended low-temperature
stage for a long time, which makes a long-lasting flat optical peak. They suggested that transition from
the state of non-optically thick winds to a state of optically
thick winds may take place. Such a transition may be accompanied by violent activities like oscillatory
behaviors in the light curves of few slow novae which are considered as some relaxation processes associated with
the transition. This kind of transition in low-mass WDs ($\le$ 0.6 M$\odot$) in close binary system, may be triggered by
the companion star through the effects of (1) spin-up by the companion motion, (2)
gravity of the companion star, and (3) drag luminosity due to frictional energy deposition. Incorporating these effects,
the authors $\textbf{(Kato \& Hachisu 2011)}$ generated two trial models adopting a parameter set of the mass of white dwarf and companion, chemical composition of the envelope and the orbit. The trial light curves successfully reproduced
the flat phase of static evolution followed by a smooth decline due to wind mass
loss. From this study the WD mass was estimated to be about 0.6 M$\odot$ for all three novae, V723 Cas, HR Del, and V5558 Sgr.

\begin{figure}
\centering
\includegraphics[bb= 0 0 414 583, width=3.3in, height=4.9in,clip]{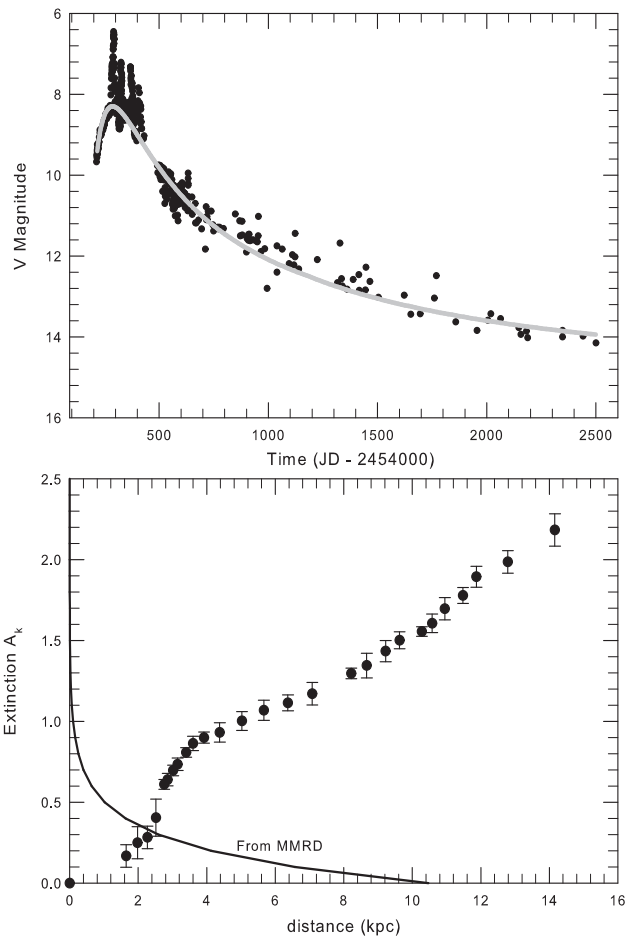}
\caption[]{The V band optical light curve of V5558 Sgr, plotted for $\sim$ 6.25 years after outburst, is shown, in the top panel.
The data are restricted to either photoelectric or CCD measurements from the AAVSO and AFOEV databases. The grey curve shows that the light curve can be
well matched by a uniform curve, obtained from a polynomial fit, in case the data points for the secondary maxima are excluded (see text for additional details).
The bottom panel shows the plot of the  extinction A$_{k}$ in the $K$ band  versus distance  towards the nova (filled dark circles; from Marshall
et al. 2006; A$_{v}$ = 11A$_{k}$). The intersection point of this extinction curve with the MMRD generated curve can be used to determine the distance
and extinction to the nova simultaneously. See text for more details.}
\label{fig2}
\end{figure}

\subsection{Distance to the nova}
If the data points of the 5 major rebrightenings or secondary maxima  are excluded, then it is found that the entire light curve can be fit by one uniform curve that initially rises and then declines. This uniform curve, on which the secondary maxima sit, is shown by the gray line  which is obtained by  a least-squares minimized  polynomial fit. From this smooth curve,  we estimate that maximum to be reached on 5 July 2007 at V = 8.3;  t$_{2}$ and t$_{3}$, the times to decline by 2 and 3 magnitudes from visual maximum, to be 281 $\pm$ 3 and  473 $\pm$ 3 days respectively. V5558 is therefore an unusually slow nova; possibly one of the slowest ever known (from the selection of well-observed novae by Warner(1995), only DO Aql has larger t$_{2}$ and t$_{3}$ values of 430 and 900d respectively). The validity of  using  maximum magnitude - rate of decline (MMRD) relations for such extremely slow novae is debatable  because most MMRD  relations are obtained as outcomes of  fits of M$_{v}$ to t$_{2}$ or t$_{3}$ values in which the latter have much smaller values.
For example, considering two well known relations, in the sample used by  della Valle $\&$ Livio (1995) the largest value of t$_{2}$ is 117d and in the Downes and Duerbeck (2000) MMRD sample there is only one nova with  t$_{2}$ greater than 100 days (viz. HR Del with t$_{2}$ = 172d).  In general not many very slow novae are known; Capaccioli et al. (1990) cite the case of one nova in the LMC with  t$_{2}$ = 117d
while Darnley et al (2006) give instances of four  novae with t$_{2}$ exceeding 100d viz. 817.2, 198, 100.3 and 331d respectively.

We thus estimate M$_{v}$ and the distance to the nova using a different approach. The variation of the extinction with distance towards the nova is shown, for a ten arc minute field around the object, in the bottom panel of  Figure 2 based on  modeling of the galactic extinction by Marshall et al. (2006). Two values of the excess $E(B-V)$ have been estimated for V5558 Sgr using independent approaches. From the interstellar NaI D1 and D2 lines, Munari et al (2007a) obtain $E(B-V)$ = 0.36 while Rudy et al (2007) obtain   $E(B-V)$ $\sim$ 0.8 based on the relative strengths of the O I 8446 and 12187{\AA}~ lines. Both methods are known to yield reliable estimates of the reddening. Using both these values of $E(B-V)$, and the relations A$_{v}$ = 3.1$E(B-V)$ and A$_{v}$ = 11A$_{k}$,  the distance to the nova from Figure 2 (bottom panel) is then constrained to lie in the range 1.3  and 1.8 kpc (i.e $d$ = 1.55 $\pm$ 0.25 kpc).

\begin{figure*}
\centering
\includegraphics[bb=0 73 553 804, width=5.0in,height=5.0in,clip]{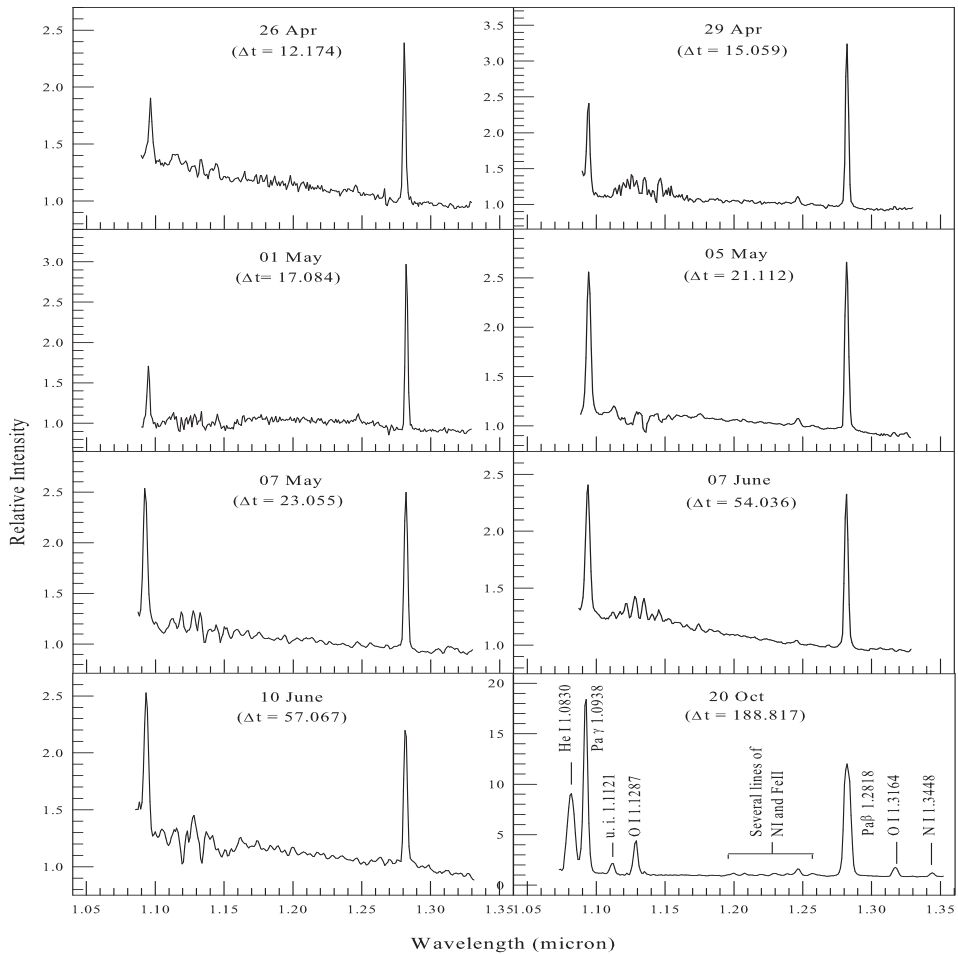}
\caption[]{The $J$ band spectra of V5558 Sgr on different days with the flux normalized to unity at 1.25 ${\rm{\mu}}$m. The prominent lines are identified and marked (see Table 3 for details; u.i = unidentified)
}
\label{fig2}
\end{figure*}

\begin{figure*}
\centering
\includegraphics[bb=0 0 559 725,width=5.0in,height=5.0in,clip]{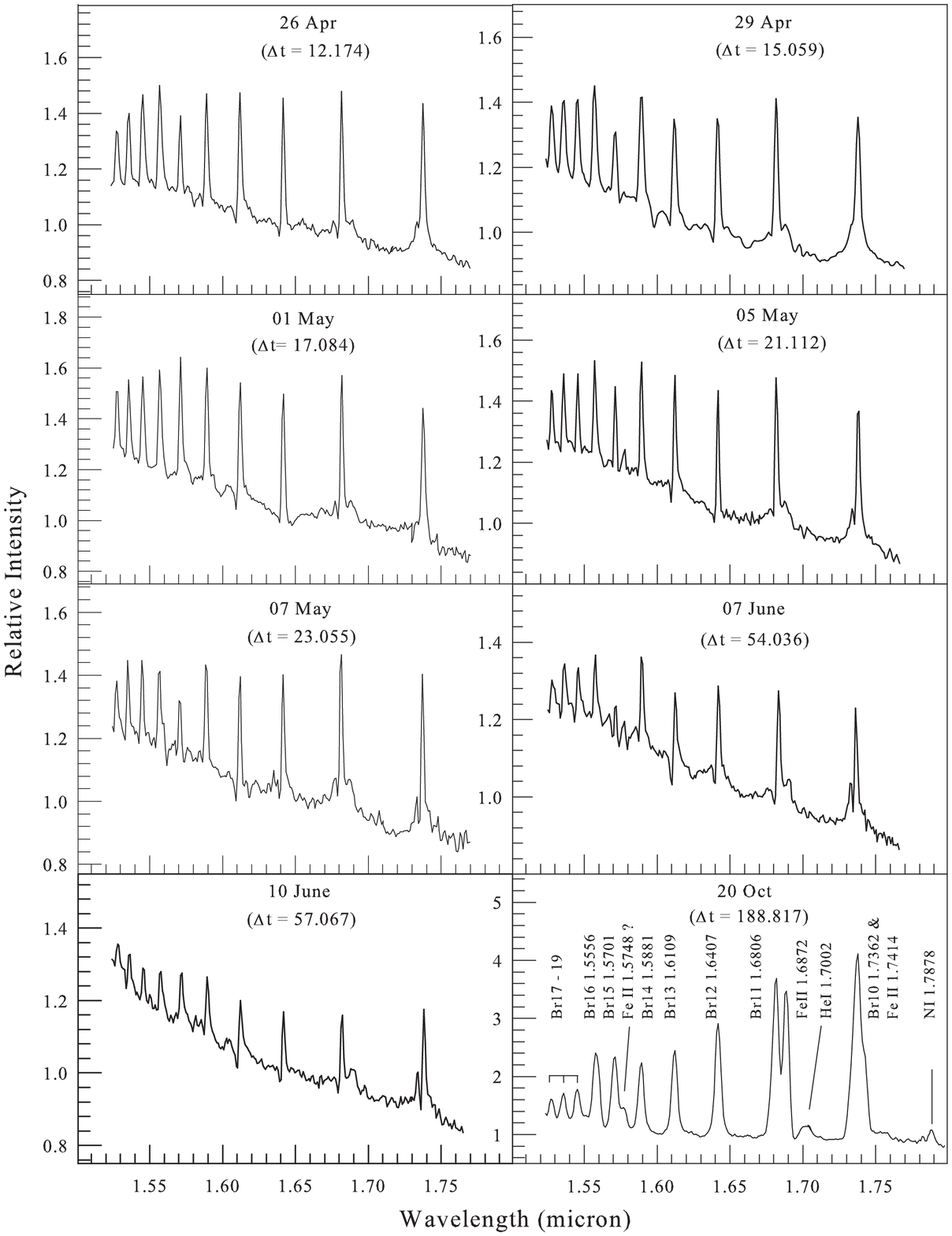}
\caption[]{ The $H$ band spectra of V5558 Sgr on different days with the flux normalized to unity at 1.65 ${\rm{\mu}}$m. The prominent lines are identified
and marked (see Table 3 for details.)}
\label{fig3}
\end{figure*}

\begin{figure*}
\centering
\includegraphics[bb=0 0 566 726,width=5.0in,height=5.0in,clip]{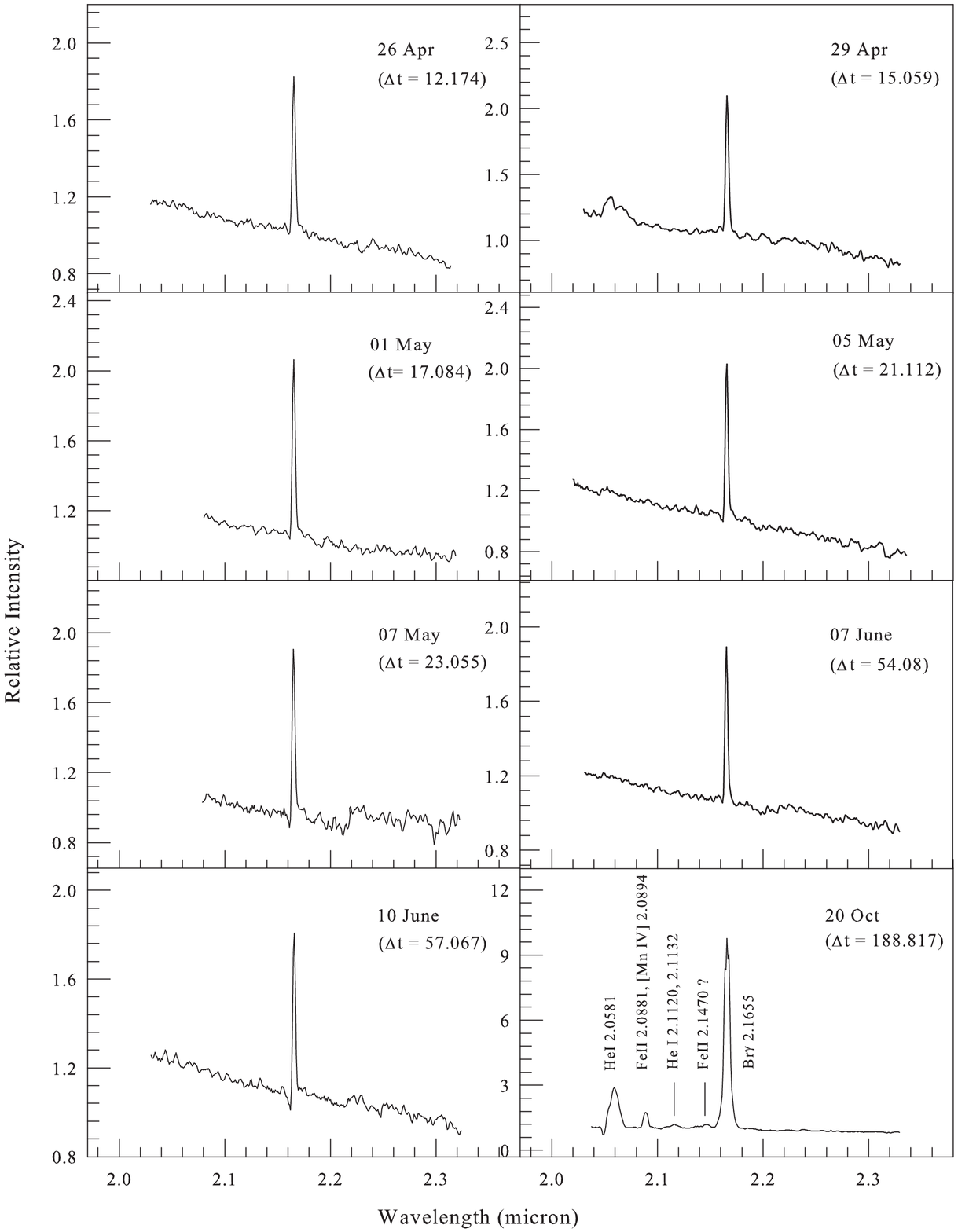}
\caption[]{The $K$ band spectra of V5558 Sgr on different days with the flux normalized to unity at 2.20 ${\rm{\mu}}$m. The prominent lines are identified
and marked (see Table 3 for details.).
}
\label{fig4}
\end{figure*}

\subsection{Line identification and general characteristics of the $JHK$ spectra, }

The near-infrared $JHK$ spectra of V5558 Sgr are presented in Figures 3, 4 and 5, respectively. The lines are marked on the figures and a list
of the identified lines are given in Table 3. The epochs of our observations are indicated by bars on the abscissa on top in Figure 1. The presented near-infrared data covers the premaxima evolution till Jun 10, our last data is on Oct 20, 11 days before the fifth maximum which was reached on 31 October. None of our spectra unfortunately coincide with the  rebrightenings to monitor spectral  changes that took place  at these secondary maxima.\\

In the initial stage, during 2007 April 26 - June 10, the $JHK$ spectra show strong emission lines of HI; these include Pa$\gamma$ and
Pa$\beta$ in the J band, Br10 to 19 in the H band and Br$\gamma$ in K band. Except the
Paschen and Brackett series lines, the only other discernible feature is the NI 1.2461, 1.2469 $\mu$m line which becomes much stronger in the later phase on 2007 Oct 20. In the $J$ band spectra OI lines at 1.1287
and 1.3164 $\mu$m which are generally seen in classically novae spectra, are weak or not present.
But it must be noted that the OI 1.1287 micron line lies in a region of poor atmospheric transmission and so it may be weakly present but not clearly discernible. An important feature
to be noted is all the emission lines are very narrow. In order to quantify the changes in the linewidths, the evolution
of Pa $\beta$ and Brackett $\gamma$ lines which are prominent but not blended with any other line, was investigated. The  full width at half maximum (FWHM) of the Pa $\beta$ and Br $\gamma$ profiles are estimated to be 533 km s$^{-1}$ and 387 km s$^{-1}$ respectively. Correcting these for instrumental broadening, our $J$ band has poorer resolution than the $K band$,  gives values of $\sim$ 240 - 260 km s$^{-1}$ for both lines indicating a very slow velocity for the ejecta.
During the pre-maxima phase, there were no significant changes seen in the width of the emission lines implying  the absence of any significant
change in the expansion velocity of the ejected envelope.
During the course of pre-maximum evolution we also note that all the emission lines
show sharp P-Cygni absorption features on the blue wings.
The values of these  absorption components for the Br 12 and Brackett $\gamma$  lines (Figure 6) are measured approximately at 530 km s$^{-1}$ from the
emission peak. \\

Significant changes in the spectra are observed on 2007 Oct 20, a few days before the fifth brightening.
The nova had entered the nebular stage at this time (Poggiani 2010; Tanaka et al. 2011).
The $JHK$ spectra show strong emission lines of Paschen and Brackett series, HeI, NI, OI and Fe II.
Among the He I lines, those at 1.0830, 1.7002 and 2.0581 $\mu$m are
prominent.
The Ly$\alpha$ fluoresced OI line 1.1287 $\mu$m and
continuum fluoresced OI line 1.3164 $\mu$m, which were absent in the pre-maximum phase, are present as normally seen in the spectra
of classical novae. The large observed ratio ($\sim$ 2.7) of the OI
1.1287/1.3164 $\mu$m lines indicates that Ly$\alpha$ fluorescence is the dominant
process contributing to the strength of the 1.1287 $\mu$m line.
In $J$ band, a magnified view of the  region between 1.19 and 1.26 $\mu$m shows a complex blend of several weak but well resolved  emission lines.
NI is clearly detected as a broad feature centered around
1.246 $\mu$m  (the NI 1.2461 and 1.2470 $\mu$m lines) on 2007 Oct 20.
Other features at 1.998, 1.2086, 1.2211, 1.2305, 1.2581 $\mu$m,
in this region are also detected which we attribute to NI lines.  Few of these features are possibly blended with adjacent weak FeII lines (see Table 3 for details).
In addition, two strong features at 1.3448 in $J$ band and 1.7878 $\mu$m in $H$ band are detected which are assigned to NI. \\

In addition to the HI, HeI, OI and NI lines, we see prominent emission features of FeII  in each of the $JHK$ bands also. These
are at 1.1126 $\mu$m which is one of the so-called
'1 micron Fe II lines' (Rudy et al. (2000)) in $J$ band, 1.6872 and 1.7414 $\mu$m in $H$ band.
All of these Fe II lines are rarely observed in the spectra
of novae, for example, in V2615 Oph (Das et al. 2009), RS Oph
(Banerjee et al. 2009,  V574 Pup (Naik et al. 2010) and V5588 Sgr (Munari et al 2014; in preparation). Apart from these, there are two more novae, namely V2540
Oph (Rudy et al. 2002) and C I Aql (Lynch et al. 2004),
where these lines appear to be detected. The excitation mechanism
for these lines is believed to be Lyman $\alpha$ and Lyman continuum fluorescence
coupled with collisional excitation (Banerjee et al. 2009
and references therein). Further, strong emissions of these Fe II lines suggest high density conditions (10$^{10}$ - 10$^{12}$ cm$^{-3}$)
inside the ejecta which is supported by recombination analysis (section 3.6) also.
As the detections of these Fe II lines in individual objects increase, it becomes amply evident  that these $H$ band lines could be present in the spectra of other novae too, but  have evaded  detection because of blending - especially when their widths are large - with the  Br 11 (1.6806 $\mu$m) line. In FeII class of novae additional blending
will take place with the strong CI line at 1.6890 $\mu$m and also CI 1.7449
$\mu$m . An instance where these FeII lines are  very clearly  resolved  is in  RS Ophiuchi (Banerjee et al. 2009), especially during the later stages of its outburst when all line widths in RS Oph narrowed due to deceleration of the ejecta.

We also detect weak emission features at 2.1140 and 2.1470 $\mu$m. The 2.1140 $\mu$m feature which was observed in V723 Cas also, is quite certainly a blend of HeI 2.1120, 2.1132 lines. The line at 2.1441 $\mu$m is probably due to Fe II. Two possible
identifications may be considered for the  emission feature  at $\sim$ 2.090 $\mu$m in the $K$ band. First, it could be FeII
2.0888 $\mu$m, a line also seen in the nova V2615 Oph (Das et al. 2009)for which an excitation mechanism by Lyman
$\alpha$ fluorescence was proposed.   Alternatively, this feature could
be the [Mn XIV] 2.0894 $\mu$m coronal line which has been seen in a few
instances in novae spectra during the coronal
phase viz., in nova V1974 Cyg (Wagner $\&$ Depoy, 1996) and in RS Oph
(Banerjee et al. 2009). However a lack of detection of optical coronal lines
at this time (Poggiani 2010) or of any other coronal lines in our spectra is not supportive of a [Mn XIV] assignment. The
present 2.0888 $\mu$m FeII line should not be confused with an
unidentified line at 2.0996 $\mu$m that has often been detected in novae and
which still remains unidentified (Rudy et al.  2002)\\

The emission lines observed on 2007 October 20 were significantly broader in comparison to the pre-maximum stage. We measured
FWHM as 1287 km s$^{-1}$ from Pa $\beta$, and 886 km s$^{-1}$ from Br $\gamma$ lines (instrumental deconvolved widths of 1201 and 880 km s$^{-1}$ respectively . While this result matches with the
findings of Munari et al. (2007b)it is puzzling why two lines from the same species should have significantly
differing velocities It is possible that they arise from different regions within the ejecta; this is known to happen in the ionized plasma discs of Be stars. No strong absorption feature associated with the emission lines is noticed.
It is also noted that no lines from low ionization species such
as Na I or MgI are seen in the $JHK$ spectra. These low ionization lines, which are indicative of low temperature conditions, have been
suggested as potential diagnostic features to predict dust formation
in the nova ejecta (Das et al. 2008). The absence of these lines in
V5558 Sgr is consistent with the lack of dust formation in this nova (Lynch et al. (2007), Rudy et al. (2007)).\\

\begin{table*}
\centering
\caption[]{List of observed lines along with fluxes in the $JHK$ spectra.}\
\smallskip
\begin{threeparttable}
\centering
\begin{tabular}{lllcccc}
\hline\
Wavelength      & Species           & Other contri-     &                & Line fluxes     &(in 10 $^{-15}$ & W cm${^{-2}}$ $\mu{^{-1}}$) \\
(${\rm{\mu}}$m) &                   & buting lines  	& Apr 29         & May 05          & Jun 10          & Oct 20 \\
\hline
\hline \\
1.0830          & He \,{\sc i}      &  	               & ...			 & ...             & ...             & 973.0 $\pm$ 27.0\\	
1.0938   	   	& Pa $\gamma$     	&                  & 32.7 $\pm$ 8.0	 & 50.7 $\pm$ 3.0  & 72.3 $\pm$ 5.0  & 1240.0 $\pm$ 30.0 \\
1.1121   		& u.i. \tnote{a} / Fe \,{\sc ii}?   &  & ...			 & ...             & ...             & 80.5 $\pm$ 5.0\\
1.1287   		& O \,{\sc i}      	& 	               & ...		     & ...             & ...             & 221.0 $\pm$ 20.0\\
1.1998          & N \,{\sc i}       &           	   & ...        	 & ...             & ...             & 21.4 $\pm$ 1.4\\	
1.2087          & N \,{\sc i}       &                  & ...             & ...             & ...             & 18.5 $\pm$ 0.5 \\
1.2211   		& N \,{\sc i}       &                  & ...             & ...             & ...             & 10.0 $\pm$ 1.0\\
1.2305          & N \,{\sc i}       &                  & ...             & ...             & ...             & 16.0 $\pm$ 0.5\\
1.2393          & u.i. \tnote{a} /Fe \,{\sc ii}?     & & ...    		 & ...	           & ...             & 7.0 $\pm$ 0.5	\\
1.2461,70       & N \,{\sc i}       &                  & 3.2 $\pm$ 0.7 	 & 3.7 $\pm$ 0.5  & 0.8 $\pm$ 0.2  & 45.0 $\pm$ 4.0\\
1.2581          & N \,{\sc i}       &                  & ...             & ...             & ...             & 17.0 $\pm$ 0.5\\
1.2818   		& Pa $\beta$        &                  & 67.8 $\pm$ 5.0	 & 57.8 $\pm$ 3.5  & 66.7 $\pm$ 4.0  & 985.0 $\pm$ 17.0\\
1.3164   		& O \,{\sc i}		&                  & ...             & ...             & ...             & 80.5 $\pm$ 2.0 \\
1.3448          & N \,{\sc i}       &                  & ...    		 & ...	           & ...             & 32.0 $\pm$ 0.5\\
1.5260   		& Br 19             &                  & 2.5 $\pm$ 0.3	 & 2.0 $\pm$ 0.4   & 1.4 $\pm$ 0.3   & 13.4 $\pm$ 2.0 \\
1.5342   		& Br 18             & 	               & 3.1 $\pm$ 0.5	 & 2.3 $\pm$ 0.3   & 1.9 $\pm$ 0.4   & 20.9 $\pm$ 2.0\\
1.5439   	   	& Br 17             & 	               & 3.1 $\pm$ 0.4	 & 2.4 $\pm$ 0.5   & 1.7 $\pm$ 0.4   & 25.2 $\pm$ 3.0 \\
1.5556   		& Br 16             &                  & 4.1 $\pm$ 0.5   & 3.5 $\pm$ 0.3   & 2.8 $\pm$ 0.3   & 78.7 $\pm$ 4.0\\
1.5701  		& Br 15        & Fe \,{\sc ii} 1.5748? & 2.3 $\pm$ 0.3   & 2.3 $\pm$ 0.4   & 1.9 $\pm$ 0.5   & 84.1$\pm$ 4.0\\
1.5881   		& Br 14             &                  & 4.4 $\pm$ 0.4   & 3.2 $\pm$ 0.3   & 3.0 $\pm$ 0.4	 & 70.0 $\pm$ 3.0\\
1.6109    		& Br 13    		    &                  & 4.2 $\pm$ 0.3   & 3.6 $\pm$ 0.4   & 3.1 $\pm$ 0.3   & 91.0 $\pm$ 3.0\\
1.6407  		& Br 12          	&	               & 4.2 $\pm$ 0.3   & 3.7 $\pm$ 0.4   & 2.8 $\pm$ 0.4   & 120.0 $\pm$ 5.0\\
1.6806   		& Br 11    		& Fe \,{\sc ii} 1.6872 & 4.9 $\pm$ 0.3   & 5.4 $\pm$ 0.6   & 4.2 $\pm$ 0.5   & 166.6 $\pm$ 5.2\\
1.6872          & Fe \,{\sc ii} &   	               & 0.4 $\pm$ 0.1 & 0.5 $\pm$ 0.1  & 0.8 $\pm$ 0.3  & 142.0 $\pm$ 5.5\\
1.7002       	& He \,{\sc i}  &                       & ... 	         & ...             & ...             & 28.8 $\pm$ 2.0\\
1.7362  		& Br 10         & Fe \,{\sc ii} 1.7414  & 6.4 $\pm$ 0.3  & 6.1 $\pm$ 0.5   & 5.5 $\pm$ 0.4   & 205.1 $\pm$ 4.8\\
1.7878          & N \,{\sc i}   &                       & ...            & ...             & ...             & 16.7 $\pm$2.5 \\
2.0581 			& He \,{\sc i}  &                       & ... 			 & ...             & ...             & 420.0 $\pm$ 6.0\\
2.0888			& Fe \,{\sc ii} & [MnXIV] 2.089?	    & ...            & ...             & ...             & 26.5 $\pm$ 0.2 \\
2.1120,32       & He \,{\sc ii} &                       & ...            & ...             & ...             & 14.0 $\pm$ 0.2 \\
2.1470          & Fe \,{\sc ii} & 	                    & ...            & ...             & ...             & 7.6 $\pm$ 1.0\\
2.1655   		& Br $\gamma$   &                     	& 6.8 $\pm$ 0.3	 & 7.6 $\pm$ 0.4   & 7.2 $\pm$ 0.4   & 410.0 $\pm$ 6.0\\
\hline
\end{tabular}
\begin{tablenotes}
\item[a] u.i. = unidentified
\end{tablenotes}
\end{threeparttable}
\end{table*}

\subsection{Spectral classification}

Williams (1992) introduced two principal classes for novae, the "Fe II" and "He/N" classes, based on the  strength of of non-hydrogen emission lines in the optical spectra  in the early stage.
In addition, a small fraction of novae display characteristics of both classes and these are referred to as "hybrid" or "Fe IIb" novae. These novae have strong Fe II lines soon after the outburst which are broader than those seen in typical "Fe II" novae and subsequently display "He/N" spectrum. An example of "hybrid" nova is V574 Pup (Naik et al. 2010; also see Williams 1992). The optical classification scheme has been extended into near-Infrared spectral region $JHK$ (1.08 to 2.35 $\mu$m) by Banerjee $\&$ Ashok (2012). In the near-infrared, the principal differentiating feature between the two classes is the strong presence of carbon lines in "Fe II" novae and their absence in He/N novae. The strongest of these C lines appear in the J band at 1.166 and 1.175 $\mu$m and in the H band at 1.6890 $\mu$m and several lines between 1.72 to 1.79 $\mu$m. On the other hand, He/N novae show prominent He emission right from the beginning of the outburst. The principal He lines are generally seen at 1.0830, 1.7002, 2.0581 and the 2.1120,2.1132 $\mu$m. Few  near-infrared lines of nitrogen are detected in both Fe II or He/N spectra (for e.g the 1.2461, 1.2469 $\mu$m lines) but they are invariably stronger in He/N novae as observed in optical spectra also (Williams 1992).\\

V5558 Sgr comes closest to being classified as a hybrid nova but with an anomalous component. The hybrid tag is justified because it showed a transition from one class to another but the anomaly arises because the transition was in the reverse direction viz from He/N to FeII instead of the other way around. Only one other hybrid nova T Pyx has shown such a similar reverse transition, both in the optical and near-infrared (see Joshi et al. 2014 for the  near-infrared  behavior and references therein for the optical results). Let us review the  optical and  near-infrared  behavior of V5558 Sgr and compare them using the present results and those from Tanaka et al (2011) because they share common dates (or very nearby) of observation between the two.

In the optical, on April 16  it appeared  as a He/N type nova but by 2007 April 25  Fe II lines and a  Fe II class   spectrum appeared showing that it had  evolved from the He/N type to  the Fe II type. T Pyx made a similar transition within a week or less (Joshi et al. 2014).  In the near-infrared, there is no spectrum on 16 April to compare with the optical. But for the Tanaka et al (2011) spectrum of 25 April in the optical, the  near-infrared spectra of 26 April is available and this is certainly not of the Fe II type.  No prominent CI lines are present as was clearly seen in T Pyx. The near-infrared spectra at this stage more closely resemble the He/N class (there is a caveat attached to this sentence which we qualify later). All subsequent optical spectra between 25 April to    10 June are of typical of the Fe II class but surprisingly all the  near-infrared spectra during this same time are more close to the He/N type. This is a major difference in the optical and $\textbf{infrared}$ behavior and a significant finding from this work. This adds complexity to any unification scheme (Williams 2012; Shore 2013) that tries to explain Fe II, He/N and hybrid novae simultaneously.

In the paradigm proposed by Williams (2012),  the  the He/N spectra have their origin in the material lost from the white dwarf surface during the thermonuclear runaway. That is why the He lines,    responsible jointly with N lines for the creation of the He/N tag, are only seen immediately after the outburst before the He-rich nova ejecta material intermingles with  a large circumbinary envelope of gas whose origin is the secondary star. The dominance of the He lines is thus decreased. The  Fe II spectra point to their formation in this large circumbinary envelope of gas which is proposed to be created by the impact of the nova ejecta with the secondary star. Depending on the proximity of the secondary star, the WD ejecta mass, and the radiation  field impinging on the secondary star, the secondary star could be stimulated to eject mass. Hybrid objects are explained by changing parameters in the two emitting regions (one comprising of the WD ejecta material and the other of the circumbinary envelope) during the post outburst decline.  How the contrasting IR behavior (vis-a-vis the optical behavior) in V5558 Sgr could be reconciled within this scheme will be interesting to see.  We do not have  a easy or straightforward answer to explain  the observed behavior but are working on a general model that tries to reproduce both the Fe II and He/N type spectra.  In such a model, we feel there is a need for including an additional parameter which considers the changing excitation and ionization conditions in the ejecta resulting from  changes in the central WD's temperature.

If we do not take help of the  optical spectra and base a classification  solely on just the near-infrared spectra during and around maximum (i.e. spectra between April to June), the caveats associated with a He/N classification throughout this period are as follows. First the He lines, the important ones being expected  at 1.0830, 1.7002 and 2.0581,  are very weak or missing. However NI is present in the form of NI  1.2461, 1.2469 $\mu$m right from the earliest stage at a strength roughly consistent with that expected in the He/N class. But there is another troubling anomaly. The observed width of the lines are very narrow whereas He/N novae invariably have line widths of several thousands of km/s. Thus, on the whole, the spectra of V5558 Sgr emphasizes both the complexity and the diversity of novae spectra.

\begin{figure}
\centering
\includegraphics[bb=95 60 500 350,width=3.3in,height=2.75in,clip]{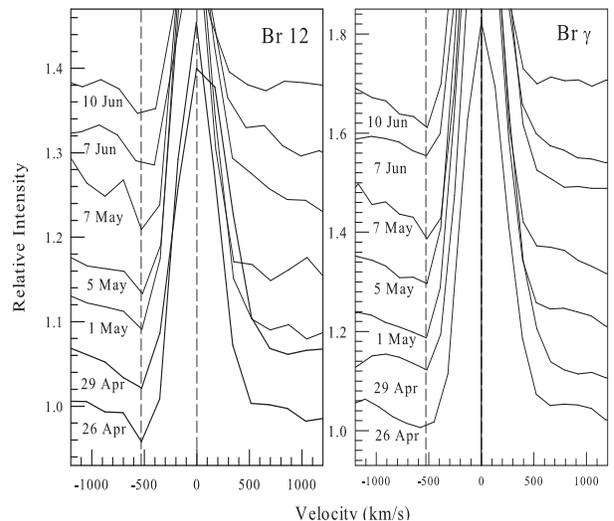}
\caption[]{The time evolution of the Br12 and Br $\gamma$ 2.1655 $\mu$m lines showing
the P-Cygni profile in the early stages with its absorption feature at $\sim$ 530 km s$^{-1}$ blueward of the emission peak.}
\label{fig1}
\end{figure}

\subsection{Recombination analysis}
A recombination case B analysis of the HI line strengths was performed
  for the spectra of all days. Representative results for four days viz. 2007 April 29, May 05, Jun 10 and Oct 20 are presented in
Figure 7 which represents the observed strength of the Brackett lines with
respect to Br14 normalized to unity. Debelending of the H lines with contaminating lines (for e.g. Br 10 and Br 11 with FeII lines) was done either manually (see Figure 8 as an example) or using IRAF tasks. The
observed H line strengths, presented in Table 3, were then compared with that expected by using case B emissivities from Storey $\&$ Hummer (1995). From
recombination theory it is qualitatively expected, under optically thin conditions, that when strengths of lines of
the same series are compared, a lower line of the series should be stronger than a higher line. For example, it is
expected that Br$\gamma$ (corresponding to a transition between levels 7--4) is expected to be significantly stronger
than any higher line of the series like Br10 or Br11 (transitions between 10 -- 4 and 11 -- 4 respectively).
But the reverse is actually being observed here on all epochs except for 2007 20 October. This clearly indicates that at these epochs  the Br $\gamma$ line is  optically thick and possibly so are the other Br lines also. The comparison has been made here with a representative Case B model at T = 10000K and density = 10$^{10}$ cm$^{-3}$ but our conclusions will remain equally valid for  case B values over an extended
parameter space of electron density n$_{e}$ varying between 10$^{5}$ and 10$^{13}$ cm$^{-3}$
and temperature varying between 5000 and 20000 K. This may be seen for e.g. in a similar  detailed graphical illustration done for the novae RS Oph $\textbf {(Figure 6, Banerjee et al. 2009)}$ and T Pyx $\textbf{(Figure 9, Joshi et al. 2014)}$.

Because the Br lines  appear optically thick, their  actual strengths are  unknown and hence it is not possible
to estimate the mass of the nova by comparing observed and predicted case B line luminosities. However, the emission measure ${n_e}^2L$ can be constrained where  $L$ is the linear size of the emitting
region. The values of the opacity  ${\Omega}_{n,n'}$ for different H{\sc i}
lines for transitions between levels $(n,n')$ at different temperatures and
densities are known (tabulated in Hummer $\&$ Storey 1987; Storey $\&$ Hummer 1995). The optical depth at line-center ${\tau}_{n,n'}$ can be calculated using
$\tau$ = ${n_e}{n_i}{\Omega}L$, where $L$ is the path length in cm.
From Figure 8, among the Brackett lines, it can be clearly seen the Br $\gamma$ line is definitely optically thick. Thus we use the Br$\gamma$ line and assume with considerable certainty that its optical depth is greater
than unity i.e. $\tau$(Br$\gamma$)$>$ 1.

From Storey $\&$ Hummer (1995) we note that the value of the opacity factor ${\Omega}_{n,n'}$ does not vary too much with
density or temperature for the near-infared H{\sc i} lines including Br$\gamma$ ( the quantitative change is mentioned a sentence later). Since in the
early outburst stages, the electron density in the nova ejecta is expected to be high (values as high as $n_e$ = 10$^{14}$ cm$^{-3}$ are
even invoked) we make a conservative estimate that $n_e$ is in the range 10$^{9}$ to 10$^{13}$ cm$^{-3}$.
 Using the range of values of ${\Omega}_{n,n'}$ between 1.3 $\times$ 10$^{-34}$ to 7.46 $\times$ 10$^{-34}$ corresponding to
$n_e$ values varying between 10$^{9}$ to 10$^{13}$ cm$^{-3}$  for a representative
temperature $T_e$ = 10$^4$ K, we get the emission measure ${n_e}^2 L$ to be in the
range 1.3 $\times$ 10$^{33}$ to 7.7 $\times$ 10$^{33}$ cm$^{-5}$.

Approximate constraints on the electron density can be obtained
if we can take $L$ to be the kinematic distance $v {\times} t$ traveled by the
ejecta where $v$ is the velocity of ejecta and
$t$ is time after outburst. Using a representative value of $v$ = 250 kms$^{-1}$ as discussed earlier and $t$
ranging from 12 to 188 days, and applying the constraint that
$\tau$(Br$\gamma$)= ${n_e}{n_i}{\Omega}L$ $>$ 1; we obtain lower limits for
 $n_e$ in the range 1.7 $\times$ 10$^{9}$ cm$^{-3}$ to 1.6 $\times$ 10$^{10}$ cm$^{-3}$ (assuming $n_e$ = $n_i$). These
 lower limits are likely to be smaller than the actual $n_e$ values because $\tau$(Br$\gamma$) can be
 considerably $>$ 1 that has been used above. The density in the ejecta is seen to be fairly high over the duration of our observations. It is worth noting Lynch et al. (2000) show
 that high
 densities of 10$^{10}$ cm$^{-3}$ or more tend to thermalize the level populations through
 collisions and thereby bring about deviations from Case B predictions  as is observed here.

By 20 October, the ejecta appears to have diluted sufficiently to be optically thin and the
 line strengths follow a caseB distribution  well. We estimate of the mass of the emitting gas at this epoch  using the same approach as in some earlier novae by using the following relation (Banerjee et al. 2010, Raj et al. 2013):

\begin{equation}
M = (4 \pi d{^ {\rm 2}} (m{_H}){^2}(f V/ \epsilon)){^ {\rm 0.5}}
\end{equation}

where $d$ is the distance, $m{_H}$ the proton mass, $f$ the observed flux
in a particular line, $\epsilon$ the corresponding case B emissivity; $V$ is the
volume of the emitting gas which equals $[4/3 \pi (vt){^3} \phi]$, where $\phi$, $v$ and
$t$ are the filling factor, velocity and time after outburst, respectively.
We assume T = 10000K and n$_{e}$ to be $\sim$ 10$^{10}$ cm$^{-3}$ prompted both by the goodness
of the fit in the bottom panel of Figure 7 and for the following additional reason.
As discussed above, n$_{e}$  was  close to $\sim$  10$^{10}$ cm$^{-3}$ during our June observation (57d after outburst) but most likely had an even higher value. If the ejecta subsequently diluted geometrically (n$_{e}$ $\propto$ r$^{-2}$) thereafter upto 20 October (188d after outburst), the density would drop by a factor of 10 (i.e. square of 57/188).
Based on this choice of n$_{e}$, we estimate the mass of the gas in the ejecta in the range (6.0 $\pm$ 1.5) $\times$ 10$^{-4}$ M$_{\odot}$. This is a standard value for novae. However, this mass estimate is restricted in accuracy by our lack of knowledge of the filling factor which is assumed to be  1 here .

\begin{figure}
\centering
\includegraphics[bb=0 312 294 711, width=2.75in,height=4.0in,clip]{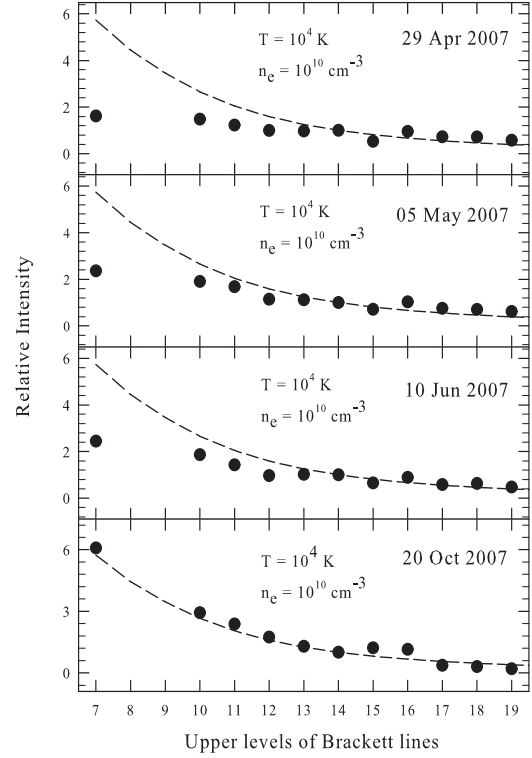}
\caption[]{Recombination analysis for the hydrogen Brackett lines in V5558 Sgr
on selected dates. The abscissa is the upper level number of the Brackett
series line transition (Br7 = Br$\gamma$). The line intensities (filled circles) are relative to that of Br 14. The
Case B model predictions for the line strengths are also shown by dashed lines.}
\label{fig1}
\end{figure}


\begin{figure}
\centering
\includegraphics[bb= 0 0 450 295, width=3.30in,height=3.0in,clip]{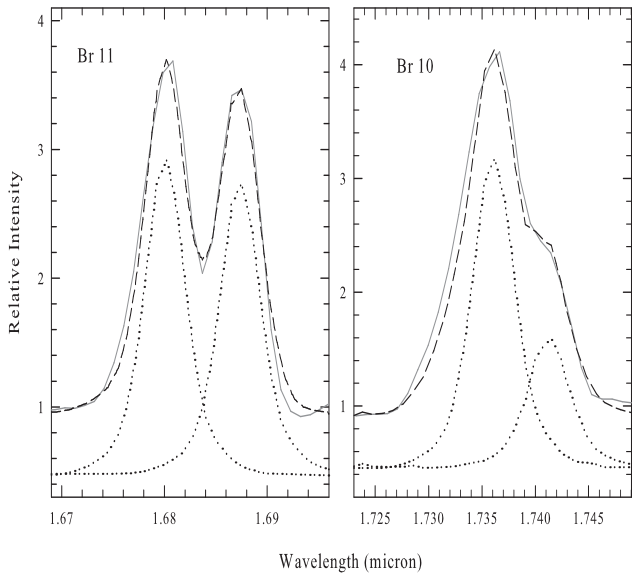}
\caption[]{Deblending of the 2007 Oct 20  profiles of the  Br 11 (1.6806 $\mu$m) and  B10 (1.7362 $\mu$m) lines with neighboring   Fe II 1.6872 and 1.7414 $\mu$m lines respectively using two-gaussian fits. The individual gaussians are shown by the dotted lines, their co-added sum by the dashed lines and the observed data by the gray lines.}
\label{fig1}
\end{figure}

\begin{figure}
\centering
\includegraphics[bb= 0 5 320 572, width=2.75in,height=4.5in,clip]{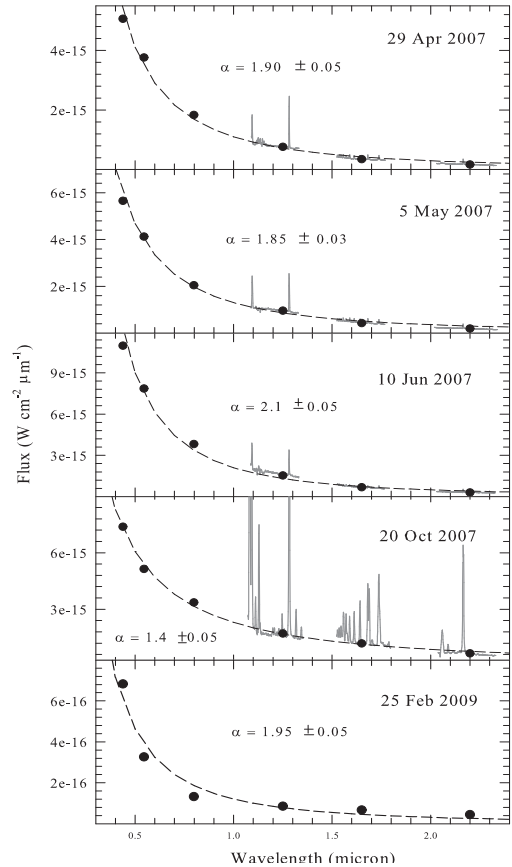}
\caption[]{ The SED for nova V5558 Sgr is shown using $BVIJHK$ fluxes for five representative dates 2007 April 29, 2007 May 5, 2007 June 10, 2007 October 20
and 2009 February 25 are shown. $BVI$ fluxes are from AAVSO, $JHK$ fluxes on first four days
are from Mount Abu observations. $JHK$ magnitudes on 2009 February 25 are collected from SMART observations. The composite $JHK$ spectra for the first
four representative dates are shown also. Model fits to the data with a power law, F$_{\lambda}$ $\propto$ $\lambda$$^{-\alpha}$, are shown by the
continuous lines with the broad-band fluxes represented as filled circles.}
\label{fig1}
\end{figure}


\subsection{Evolution of the continuum}

We analyze the evolution of the continuum flux distribution to check for any sign of dust formation and also analyze the general trend in the temporal evolution of the spectral energy distribution (SED). The continuum has been modeled with power-law
fits ( $F_{\lambda} = \lambda^{-\alpha}$), the results of which are shown in Figure 9, wherein
representative spectra on 2007 April 29, May 5, June 10 and October 20, sampling the duration of our observations have been plotted.
The spectra have been flux calibrated using contemporaneous $JHK$ values from Table 2 or  data from Walter et al. (2012) after correcting these for reddening using E(B - V) = 0.58
as the average of the values provided by Munari et al. (2007a) and Rudy et al. (2007).
In the early stage i.e. during April 26 - May 5 of the evolution of V5558 Sgr, the continuum spectrum
is well matched with fits having spectral indices of $\alpha$ $\sim$ 1.85 - 1.9.
The subsequent spectra, observed during Jun 7 - 10, tend to become slightly steeper with the
slope of $\alpha$ = 2.1. Around the fifth maximum the SED becomes flatter with $\alpha$ = 1.4 $\pm$
0.5. Proceeding further, we have also specifically tried to fit the SED
for 2009 February 25 which is about 317 days after the discovery. The result is shown in the last panel in Figure 9 where we have
used $JHK$ data from Walter et al. (2012) and $BVI$ magnitudes from AAVSO database.
We find that the continuum is well fitted by a power law, with the spectral index 1.95 $\pm$ 0.5.\\

Such study of evolution of continuum has been performed in case of some other novae also. For example, in V574 Pup (Naik et al. 2010)
approximate value of the spectral index $\alpha$ was observed to decline from a value of 2.75 in the initial stage. In KT eridani
(Raj et al. 2013) the value of the spectral index was found to decrease from 3.34 to 2.67.
However, in the case of V4633 Sgr (Lynch et al. 2001) the change in the slope of the continuum was
found to be in the opposite direction. The slope changed from 2
to 2.7 in observations taken 525 and 850 days after outburst.
In comparison, the behavior of V5558 Sgr is peculiar where $\alpha$ did not increase or decrease monotonically as observed in other novae. This appears to be a consequence of the very slow evolution of the nova. It may be noted that a blackbody, in the Rayleigh-Jeans regime, is expected to give a index of 4, quite different from that seen here.\\

No evidence for dust formation is seen, which in  case it is formed, will show up as a significant excess at the infrared wavelengths. This is not the case here. Observations by the Wide field Infrared Survey Explorer (WISE) too, at an even later date on 2010 23 March, do not show any source with infrared excess at the position of the nova indicating the non-formation of dust.\\

\section{Summary}

We have presented multi-epoch near-infrared spectroscopy and photometry observations of the slow nova V5558 Sgr which was discovered in April 2007
and which showed a sequence of multiple secondary outbursts.
The optical light curve has been analyzed allowing a distance estimate of  1.55 $\pm$ 0.25 kpc to
be made.  V5558 Sgr is a fairly unique nova by virtue of showing a  rare reverse-hybrid transition
from He/N to Fe II type, based on its optical spectra.
However the near-infrared data do not show such a transition and a discussion is made of this aspect. A Case B recombination analysis of
the hydrogen lines has been done allowing  the mass  of the gaseous component in the ejecta to be estimated to be   (6.0 $\pm$ 1.5) $\times$ 10$^{-4}$
M$_{\odot}$ assuming a filling factor of unity.

\section{Acknowledgments}
The research work at S N Bose National Centre for Basic Sciences is funded by the Department
of Space and Technology, Government of India. The research work at Physical Research Laboratory  is funded by the Department
of Space, Government of India.  We are thankful to AAVSO, USA and AFOEV, France for the use of their optical photometric data.


\begin{thebibliography}{99}

\bibitem[\protect\citeauthoryear{Banerjee}{2003}]{b1}Banerjee D. P. K., Ashok N. M., 2012, BASI, 40, 243


\bibitem[\protect\citeauthoryear{Banerjee}{2002}]{b3} Banerjee D. P. K.,  Ashok N. M., 2002, A$\&$A, 395, 161

\bibitem[\protect\citeauthoryear{Banerjee}{2009}]{b4} Banerjee D. P. K., Das R. K., Ashok N. M., 2009, MNRAS, 399, 357

\bibitem[\protect\citeauthoryear{Banerjee}{2010}]{b5} Banerjee D. P. K., Das R. K., Ashok N. M. et al., 2010, MNRAS, 408, L71


{\bibitem[\protect\citeauthoryear{Capaccioli}{2008}]{b8} Capaccioli M., Della Valle M., D'Onofrio M. \& Rosino L., 1990, 360, 63}

\bibitem[\protect\citeauthoryear{Darnley}{2006}]{b8} Darnley M. J., Bode M. F., Kerins E. et al., 2006, MNRAS, 369, 257

\bibitem[\protect\citeauthoryear{Das}{2008}]{b8} Das R.K., Banerjee D. P. K., Ashok N. M., Chesneau O.,  2008, MNRAS, 391, 1874

\bibitem[\protect\citeauthoryear{Das}{2009}]{b7} Das R.K., Banerjee D.P.K., Ashok N.M.,  2009, MNRAS, 398, 375

\bibitem[\protect\citeauthoryear{della Valle}{1995}]{b8} della Valle M., Livio M., 1995, ApJ, 452, 704

\bibitem[\protect\citeauthoryear{Downes}{2000}]{b10} Downess R. A., Duerbeck H. W., 2000, AJ, 120, 2007

\bibitem[\protect\citeauthoryear{Evans}{2003}]{b9} Evans A., Gehrz R. D., Geballe T. R. et al., 2003, AJ, 126, 1981

\bibitem[\protect\citeauthoryear{Friedjung}{1992}]{b10} Friedjung M., 1992, A\&A, 262, 487


\bibitem[\protect\citeauthoryear{Henden}{2007}]{b12} Henden A., Munari U., 2007, IBVS, 5803


\bibitem[\protect\citeauthoryear{Hummer}{1987}]{b13} Hummer D. G., Storey P. J., 1987, MNRAS, 224, 801.

\bibitem[\protect\citeauthoryear{Iiijima}{2007a}]{b18} Iijima T., 2007a, CBET, 934

\bibitem[\protect\citeauthoryear{Iiijima}{2007b}]{b17} Iijima T., Correia A. P., Hornoch K., Carvajal J., 2007b, CBET, 1006

\bibitem[\protect\citeauthoryear{Joshi}{2014}]{b18}Joshi V., Banerjee D. P. K., Ashok N. M., 2014, MNRAS, 443, 559

\bibitem[\protect\citeauthoryear{Kato}{2002}]{b14} Kato T., Uemura M., Haseda K., Yamaoka H., Takamizawa K., Fujii M., Kiyota S., 2002, PASJ, 54, 1009

\bibitem[\protect\citeauthoryear{Kato}{2009}]{b15} Kato, M., Hachisu, I. 2009, ApJ, 699, 1293

\bibitem[\protect\citeauthoryear{Kato}{2011}]{b16} Kato, M., Hachisu, I. 2011, ApJ, 743, 157


\bibitem[\protect\citeauthoryear{Lynch}{2004}]{b20} Lynch D.K., Wilson J. C., Rudy R. J., Venturini C. C., Mazuk S., Miler N. A., Puetter R. C., 2004, AJ, 127, 1089

\bibitem[\protect\citeauthoryear{Lynch}{2007}]{b21} Lynch D.K., Russell R. W., Rudy R.J., Pearson R., Woodward C. E., 2007, IAU Circ. 8874

\bibitem[\protect\citeauthoryear{Marshall}{2006}]{b21} Marshall D. J., Robin A. C., Reyle C., Schultheis M., Picaud S., 2006, A\&A, 453, 635

\bibitem[\protect\citeauthoryear{Munari}{2007a}]{b22} Munari, U.; Siviero A., Dallaporta S. et al., 2007a, CBET, 965.

\bibitem[\protect\citeauthoryear{Munari}{2007b}]{b23} Munari U., Orio M., Valentini M. et al., 2007b, CBET, 1010.

\bibitem[\protect\citeauthoryear{Munari}{2014}]{b23} Munari U., Henden A., Banerjee D.P.K., Ashok N.M. , Righetti G.L., Dallaporta S. and Cetrulo G., 2014 (submitted)

\bibitem[\protect\citeauthoryear{Nakano}{2007}]{b24} Nakano S., Sakurai Y., Itagaki K., Koff R., 2007, IAU Circ. 8832

\bibitem[\protect\citeauthoryear{Naik}{2010}]{b25} Naik S., Banerjee D. P. K., Ashok N. M., Das R. K., 2010, MNRAS, 404, 367


\bibitem[\protect\citeauthoryear{Naito}{2007}]{b26} Naito H., Matsuda K., Yamaoka H., 2007, CBET 934


\bibitem[\protect\citeauthoryear{Poggiani}{2008}]{b28} Poggiani R., 2008, NewA, 13, 557

\bibitem[\protect\citeauthoryear{Poggiani}{2010}]{b29} Poggiani R., 2010, NewA, 15, 657

\bibitem[\protect\citeauthoryear{Raj}{2013}]{b30} Raj A., Banerjee D. P. K., Ashok N. M., 2013, MNRAS, 433, 2657

\bibitem[\protect\citeauthoryear{Rudy}{2000}]{b32} Rudy R. J., Mazuk S., Puetter R. C., Hamann F., 2000, ApJ, 539, 166

\bibitem[\protect\citeauthoryear{Rudy}{2007}]{b31} Rudy R. J., Lynch D. K., Russell R. W., Woodward C. E., 2007, IAU Circ., 8884

\bibitem[\protect\citeauthoryear{Rudy}{2002}]{b33} Rudy R. J., Lynch D. K., Mazuk S., Venturini C. C., Puetter R. C., perry R. B., 2002, BAAS, 34, 1162


\bibitem[\protect\citeauthoryear{Shen}{2009}]{b35} Shen K. J., Idan I., Bildsten L., 2009, ApJ, 705, 693

\bibitem[\protect\citeauthoryear{Shore}{2013}]{b35} Shore S. N., 2013, A\&A, 559, L7

\bibitem[\protect\citeauthoryear{Storey}{1995}]{b37} Storey P. J., Hummer D. G., 1995, MNRAS, 292, 41

\bibitem[\protect\citeauthoryear{Strope}{2010}]{b36} Strope R. J., Schaefer B. E., Henden A. A., 2010, AJ, 140, 34

\bibitem[\protect\citeauthoryear{Tanaka}{2011}]{b38} Tanaka J., Nogami D., Fujii M, Ayani K., Kato T., Maehara H., Kiyota S., Nakajima K., 2011, PASJ, 63, 911.

\bibitem[\protect\citeauthoryear{Terzan}{1970}]{b39} Terzan A., 1970, POHP, 10, 167.

\bibitem[\protect\citeauthoryear{Wagner}{1996}]{b40} Wagner R. M., Depoy D. L., 1996, 467, 860

\bibitem[\protect\citeauthoryear{Walter}{2012}]{b40} Walter F. M., Battisti A., Towers S. E., Bond H. E., Stringfellow G. S., 2012, PASP, 124, 1057.

\bibitem[\protect\citeauthoryear{Warner}{2008}]{b41}Warner, B., 1995, {\it Cataclysmic Variable Stars. Cambridge Astrophysics Series}, Cambridge Univ. Press, Cambridge, New York, p. 260

\bibitem[\protect\citeauthoryear{Williams}{2008}]{b42}Williams R.E., 1992, AJ, 104, 725

\bibitem[\protect\citeauthoryear{Williams}{1992}]{b43} Williams R.E., 2012, AJ, 144, 98

\end{thebibliography}
\end{document}